%% file: letter.tex
\documentclass[]{aastex701}
\usepackage{graphicx}
\usepackage{multirow}
\usepackage{amsmath}
\usepackage{subcaption}
\usepackage{CJKutf8}
\input{bangla_commands}

\begin{document}


\title{Model Predictive Scoring Shows Specific BAO Observations (not SNIa) Drives $w_0w_a$ Tension}

\author[orcid=0000-0002-5652-8870,sname='Karim',gname='Tanveer']{Tanveer Karim ({\bng tanviir kirm})}
\affiliation{David A. Dunlap Department of Astronomy \& Astrophysics, University of Toronto, 50 St. George St, Toronto, ON M5S 3H4, Canada}
\affiliation{Dunlap Institute for Astronomy \& Astrophysics, University of Toronto, 50 St. George St, Toronto, ON M5S 3H4, Canada}
\affiliation{Center for Astronomy, Space Science and Astrophysics, Independent University, Bangladesh, Dhaka 1229, Bangladesh}
\email[show]{tanveer.karim@utoronto.ca}

\author[0000-0003-2573-9832,sname='Speagle',gname='Joshua']{Joshua S. Speagle \begin{CJK*}{UTF8}{gbsn}(沈佳士)\end{CJK*}}
\affiliation{David A. Dunlap Department of Astronomy \& Astrophysics, University of Toronto, 50 St. George St, Toronto, ON M5S 3H4, Canada}
\affiliation{Department of Statistical Sciences, University of Toronto, 9th Floor, 700 University Ave, Toronto, ON M5S 3G3, Canada}
\affiliation{Dunlap Institute for Astronomy \& Astrophysics, University of Toronto, 50 St. George St, Toronto, ON M5S 3H4, Canada}
\affiliation{Data Sciences Institute, University of Toronto, 10th Floor, 700 University Ave, Toronto, ON M7A 2S4, Canada}
\email{j.speagle@utoronto.ca}

\author[orcid=0000-0002-0965-7864,sname='Hlozek',gname='Renee']{Ren\'{e}e Hlo\v{z}ek}
\affiliation{David A. Dunlap Department of Astronomy \& Astrophysics, University of Toronto, 50 St. George St, Toronto, ON M5S 3H4, Canada}
\affiliation{Dunlap Institute for Astronomy \& Astrophysics, University of Toronto, 50 St. George St, Toronto, ON M5S 3H4, Canada}
\affiliation{Data Sciences Institute, University of Toronto, 10th Floor, 700 University Ave, Toronto, ON M7A 2S4, Canada}
\email{hlozek@dunlap.utoronto.ca}

\begin{abstract}
The recent analyses of DESI DR2 BAO, Planck CMB and various Supernovae datasets have shown preferences for evolving dark energy at various levels of significance, under frequentist model comparison tests. However, the same analysis done with a pure Bayesian model comparison test can lead to the opposite conclusion due to the impact of $w_0w_a$ priors. In contrast to these in-sample model comparison route, we approach the problem along a third axis -- which model better predicts unseen data. We use the Expected Log Predictive Density (ELPD) metric to score the predictiveness of $\Lambda$CDM and $w_0w_a$CDM models, using the leave-one-redshift-block-out cross-validation estimator. The comparison metric $\Delta$ELPD is out-of-sample, unlike $\Delta \chi^2_{\rm MAP}$ and Bayes Factor, and insensitive to prior width, unlike the Bayes factor. Aggregated, we find modest preferences for $w_0 w_a$CDM, primarily driven by the BAO. Decomposing the $\Delta$ELPD score by redshift we find that a single BAO block, LRG2 ($z = 0.706$), supplies the entire BAO preference. Interestingly, the biggest outlier point, LRG1, contributes little to the model predictive scoring because both $\Lambda$CDM and $w_0w_a$CDM fail to predict the observation by equal amount. On the other hand, all SNIa datasets' $\Delta$ELPD scores are consistent with $0$, indicating that at an out-of-distribution predictive level, the $w_0w_a$CDM tension is entirely driven by one BAO point, not SNIa. 
\end{abstract}


\section{Introduction}
\label{sec:intro}

The recent claimed preference for evolving dark energy by the DESI survey DR2 baryon acoustic oscillation (BAO) measurements, combined with the cosmic microwave background (CMB) and Type~Ia supernovae (SNIa) is between $2.8\sigma$ and $4.2\sigma$, depending on which supernova compilation is used \citep{desi-dr2-bao}. A Bayesian evidence analysis of the same data however, finds $\ln B = +0.57 \pm 0.26$ in favour of $\Lambda$CDM for DESI\,+\,CMB, and recovers a preference for $w_0w_a$CDM, \citep[where $w_0$ and $w_a$ are parameters in the Chevallier-Polarski-Linder or CPL model of dark energy,][]{chevallierpolarski,linder}  only when the DES-SN5YR sample \citep{desy5-main-paper} is included. Replacing DES-SN5YR with the recalibrated DES-Dovekie catalogue \citep{desdovekie-main-paper} removes the preference entirely \citep{ong2026-handley-desi-dr2-bayesian}. Independently, per-tracer analyses argue that a small number of DESI redshift bins carry most of the signal in both DR1 and DR2 \citep{ocolgain2026-desi-dr2, ocolgain2026-dr1-anomaly, liu2024-lrg1-lgr2-dr1-anomaly, wang2024-lrg1-lrg2-dr1}.

The preferred metrics for these analyzes are the frequentist Likelihood Ratio Test using the Maximum A Posteriori Estimator, $\Delta\chi^2_{\rm MAP}$ \citep{desi-dr2-bao}, and the Bayes Factor used in Bayesian statistics \citep{ong2026-handley-desi-dr2-bayesian}. Both quantities are what is known as \textit{in-sample} statements: every data point is used to constrain the parameters and then used \textit{again} to `score' the model. A complementary question that is asked from the data is how well a model that is fitted to a current data set \textit{predicts} future (or unseen/removed) data. We argue here that this is a more robust test for genuine model, in comparison to the more commonly used model fitting that is often prone to over-fitting due to the model flexibility from extra degrees of freedom. In this paper we can use statistical cross-validation to iteratively hold out blocks of data as the `unseen' component and test (on average) which assumed cosmological model predicts these blocks more effectively.

We quantify this predictive performance with the expected log predictive density \citep[ELPD,][]{vehtari2015-psis-loo}, the standard cross-validation score in applied statistics, estimated by Pareto-smoothed importance sampling \citep[PSIS,][]{vehtari2015-psis} and generalized to the non-factorizable Gaussian likelihoods of cosmology by \cite{burkner2018-nonfactor-elpd}. $\Delta$ELPD is \textit{out-of-sample}, unlike $\Delta\chi^2_{\rm MAP}$ or the Bayes Factor. It provides a posterior average, so is insensitive to the volume of the assumed prior (unlike the Bayes factor, see Table~\ref{tab:defns} for a comparison between metrics, the questions they answer and the properties that distinguish them). Similarly, in-sample model selection metrics cannot localize the source of model preference while $\Delta$ELPD naturally provides such decomposition mechanism to localize tensions in the data space. 

\section{What Does It Mean for a Model to Be Better?}
\label{sec:theory}
\begin{table}[htbp!]
\centering
\begin{tabular}{|l|l|c|l|}
\hline
\textbf{Metric} & \textbf{Core question} & \textbf{Scoring} & \textbf{Role of prior width} \\
\hline\hline
$\Delta\chi^2_{\rm MAP}$ & Which model fits best at its best-fitting parameters? & in-sample & weak \\
\hline
$\ln B$ & Which model better explains the data, averaged over its prior? & in-sample & strong \\
\hline
$\Delta$ELPD & Which model better predicts data it has not seen? & out-of-sample & weak \\
\hline
\end{tabular}
\caption{Three model comparison metrics and the questions they answer. The three coincide only when the additional parameters are tightly constrained relative to their priors; neither condition holds for the $w_0w_a$CDM comparison.}
\label{tab:defns}
\end{table}
Model comparison metrics are routinely treated as interchangeable measures of which model is \textit{better}, but the notion of a better model depends on the metric being used. The first two metrics in Table~\ref{tab:defns} are both in-sample, i.e., every datum informs the parameters and is then scored against them. They can nevertheless disagree, and the mechanism is Lindley's paradox \citep{lindley1957}. The paradox is described as follows: if $\mathcal{M}_0$ is nested in $\mathcal{M}_1$, a likelihood-ratio test may reject $\mathcal{M}_0$ at some significance while the Bayes factor continues to prefer it, because the \textit{evidence} averages the likelihood over the \textit{prior}. Hence $\mathcal{M}_1$ must pay for its extra parameters across their entire prior volume, whether or not the data occupy it. The DESI dark energy comparison is an instance in the textbook sense, where $\Lambda$CDM is nested in $w_0w_a$CDM, and generally no theoretical argument fixes the width of the $(w_0,w_a)$ prior. This prior dependence is a feature of the Bayes factor rather than a defect of it; a model that proposes a wide range of unrealized behavior ought to be charged for doing so. Hence this aspect of Bayesian model selection often overlooked is that the prior is an \textit{integral} part of the model being tested, a functional form of a model with two different assumed priors are treated as two different models at the level of model comparison. This implies that a Bayes factor also carries the subjectivity of the experimenter. When the prior is set by convention rather than by physics, the resultant Bayes factor inherits the convention. In Section~\ref{sec:res-prior} we show that the two prior choices of Table~\ref{tab:priors} shift in the Bayes factor of $\ln B \simeq 3$ on posteriors that are indistinguishable. Neither of the first two metrics in Table~\ref{tab:defns} tests whether a model can predict a measurement to which it was not shown, which is the main function of the $\Delta$ELPD. 

\subsection{The Expected Log Predictive Density (ELPD)}
\label{sec:elpd}

To compute the predictive score we partition the actual data into two groups, viz. $y_{i}$ and $y_{-i}$, which are the \(\{i\}\) unseen and seen \(\{-i\}\) data. We estimate the posterior assuming only the seen data and then predict the unseen data. A better-performing model will yield closer prediction to the real (currently unseen) data. The relevant density for this metric is:
\begin{equation}
p(\underbrace{y_i}_{\text{withheld}}\mid\underbrace{y_{-i}}_{\text{seen}}) =
\int \underbrace{p(y_i \mid y_{-i}, \theta)}_{\text{conditional likelihood}} \, \underbrace{p(\theta \mid y_{-i})}_{\text{posterior from seen data}}\, \mathrm{d}\theta ,
\label{eq:loo-predictive}
\end{equation}
\noindent where the left-hand side is the predictive density and the right-hand shows the same through a conditional probability relationship. Note that on the right-hand side of Equation~\ref{eq:loo-predictive}, the first term is the likelihood of the unseen data given both seen data and some value of the model parameter, while the second term is the posterior estimated using only the seen data. This second term is what makes log predictive density much less sensitive to assumed priors compared to $\ln B$. We loop over the logarithm of Equation~\ref{eq:loo-predictive} and construct an average score, known as the \textit{expected} log predictive density or ELPD. When we compare two models, we can look at the difference of their ELPD scores to identify and quantify model performance as: \(\Delta\mathrm{ELPD} = \mathrm{elpd}_{\Lambda\mathrm{CDM}}
- \mathrm{elpd}_{w_0w_a\mathrm{CDM}}\). For the models we compare, $\Delta\mathrm{ELPD} < 0$ favours the evolving dark energy $w_0w_a$CDM model over $\Lambda$CDM while $\Delta\mathrm{ELPD} \approx 0$ indicates that neither model predicts the unseen data more accurately. We adopt this convention where a negative difference favours evolving dark energy for every metric in this paper, including $\ln B$ and $\Delta\chi^2_{\rm MAP}$.

\begin{figure}
    \centering
    \includegraphics[width=0.9\linewidth]{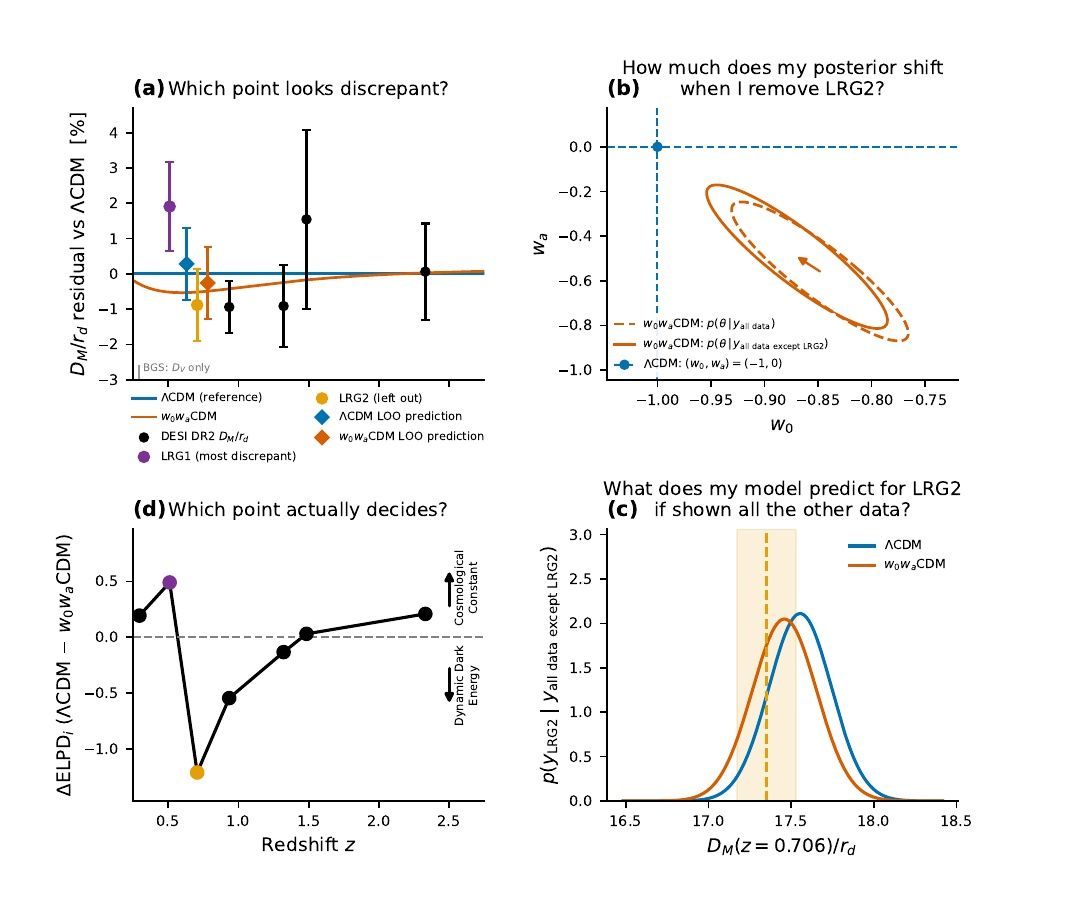}

    \caption{How leave-one-out cross-validation scores a cosmological model, at a given DESI BAO redshift. (\textit{a}) Schematic of $D_M/r_d$ data and $1\sigma$ errors shown with the best-fit models, with the withheld block highlighted in orange (LRG2, $z = 0.706$). (\textit{b}) The $w_0w_a$CDM posterior conditioned on \textit{Planck}, Pantheon+ and all seven BAO blocks (orange dashed line) and on the same data with LRG2 removed (orange solid line). The arrow connects the means and the blue crosshair denotes the $\Lambda$CDM point $(-1, 0)$. Removing LRG2 moves the posterior toward $\Lambda$CDM point, although only from a distance of $0.58$ to $0.51$, illustrating how the effect of one block on the posterior is modest even where its effect on the predictive score is decisive. (\textit{c}) The resulting leave-one-out  marginalized $D_M/r_d$ predictive density at LRG2 (Equation~\ref{eq:loo-predictive}) under $\Lambda$CDM (blue) and $w_0w_a$CDM (orange), with the measurement and its $1\sigma$ interval marked. While this panel shows marginalized posterior, the $D_M/r_d$ and $D_H/r_d$ are withheld and predicted jointly, and the score in panel (\textit{d}) uses the full two-dimensional density. Both bottom panels show that the ELPD is dominated by the DESI measurement uncertainty rather than by parameter uncertainty, as discussed in Section~\ref{sec:bao-results}. (\textit{d}) The $\Delta$ELPD per redshift block, showing that LRG2 (orange) supplies essentially all of the BAO signal, while LRG1 (purple), the largest residual in panel (\textit{a}), contributes almost none of it in the opposite `favouring $\Lambda$CDM'-direction.}
    \label{fig:pedagogy}
\end{figure}

We also note that the ELPD metric is closely related to the more well known information criteria metric such as the Akaike Information Criterion \citep[AIC,][]{aic}, Deviance Information Criterion\citep[DIC,][]{dic}, and the Widely Applicable Information Criterion \citep[WAIC,][]{waic}. These metrics may be more commonly known in astrophysics and cosmology. Along with $\Delta \chi^2_{\rm MAP}$, the DESI analysis also quotes $\Delta$DIC numbers as a way to interpret preference for dynamic dark energy. DIC has some major limitations, both from a theoretic perspective as well as practical considerations for models that are singular. A model is \textit{singular} if its \textit{best} parameter estimation projects to different values, e.g., the maximum likelihood estimator, the maximum \text{a posteriori} and the posterior mean do not agree. While is designed to be ``Bayesian", it measures the fit from a point estimate, and therefore does not give an unbiased estimation of the predictive utility of the model. Additionally, for singular models, it is unclear why the posterior mean (what DIC uses) should be the fixed point at which the model's predictive power is computed \citep{vehtari2012survey}. We estimate the ELPD scores using the statistical package \citep[\texttt{arviz},][]{arviz}, where importance weights are normalized using a generalized Pareto distribution fit to the upper tail of the distribution of the simulated importance ratios \citep[so-called Pareto-smoothed importance sampling or PSIS,][]{vehtari2015-psis} to compute the posterior term in Equation~\ref{eq:loo-predictive} from a single chain. 

\section{Measuring ELPD from Publicly Available Likelihoods}
\label{sec:methods}

To compute our scores we use the DESI DR2 baseline data combinations, namely  the DESI DR2 BAO likelihood:  \citep[\texttt{bao.desi\_dr2},][]{desi-dr2-bao}, the \textit{Planck} temperature and polarization likelihoods \citep[\texttt{planck\_2018\_lowl.EE} and \texttt{planck\_2018\_lowl.TT},][]{planck2018-lowl-TT-EE}, combined with the CamSpec high-$\ell$ TTTEEE likelihood\citep[\texttt{planck\_2018\_highl\_CamSpec2021.TTTEEE},][]{camspec-highl-2021} and CMB lensing likelihoods, namely \citep[\texttt{planck\_2018\_lensing.clik},][]{planck2018-cmblensing-clik}, and one of four Type~Ia supernova compilations, namely Pantheon+\citep[\texttt{sn.pantheonplus},][]{pantheonplus}, the Dark Energy Survey  \citep[\texttt{sn.desy5},][]{desy5-main-paper}, the recalibrated DES-Dovekie \citep[\texttt{sn.desdovekie},][]{desdovekie-main-paper}, and Union3 compilation \citep[ \texttt{sn.union3},][]{union3}. We include these likelihoods in the \texttt{cobaya} sampler \citep{cobaya2012,cobaya2019}, more details on the cosmological inference are given in Appendix~\ref{app:settings}. 

The CMB likelihood is part of $y_{\rm seen}$ at every step and is never withheld, which is a deliberate choice because the CMB constrains the late-time expansion only through a global geometric degeneracy, and hence is not a redshift-local block to withhold. We extend the idea of the classical leave-one-out cross-validation (LOOCV) estimator of ELPD to a leave-redshift-group-out estimator. Classical LOOCV removes a single data point at a time to compute the predictive score. We are, however, more interested in understanding whether a model can predict a different redshift more accurately, and hence we compute the hold-out analysis in redshift blocks. For the BAO, the natural block is the whole BAO sphere at that redshift. Every DESI sample except the Bright Galaxy Survey (BGS) in the lowest bin supplies two measurements, $D_M/r_d$ and $D_H/r_d$, giving $K = 7$ blocks comprising $13$ observables. For the supernova data, we sort by $z_{\rm cmb}$ and take contiguous equal-count blocks: $30$ blocks of $53$ for Pantheon+, $31$ of $59$ for DES-Y5, and $35$ of $52$ for DES-Dovekie. Union3 is distributed already binned in redshift and requires no further grouping ($K = 22$). We note that the redshift `width' of the blocks differ between compilations. We also consider a \textit{joint} hold-out, in which the BAO block at a given redshift is withheld together with all supernovae falling within the same redshift bin. We define the boundaries of the group as the midpoint between the BAO redshifts, which leads to lobsided cases where high-$z$ redshift groups consisting of quasar (QSO) and Ly-$\alpha$ may have no corresponding supernovae in the redshift bin. This is an artefact of the design, given that we intend to quantify how the joint model information propagates through redshift groups that contain enough supernovae. We additionally construct two low-$z$ bins, $0 \leq z < 0.1$ and $0.1 \leq z < 0.2,$ as we did not want the very low-$z$ supernovae to overwhelm the BGS BAO point in that bin. This allows us to probe the predictive power of cosmological models at very low redshifts.  The BAO-only (and the supernovae-only) hold-outs quantify the predictive power of the model given one tracer to predict another tracer's measurement at the same epoch, using information from the complementary tracer in the same redshift group to adjust the posterior. The joint hold-out instead predicts the entire expansion history at that epoch, with no probes available to support the model predictive power within an epoch.

\section{Results}
\label{sec:results}

\begin{table*}[htbp!]
\centering
\begin{tabular}{lrrrrrr}
\hline\hline
 & & & & \multicolumn{2}{c}{Separate hold-out} & Joint hold-out \\
\cline{5-6}\cline{7-7}
SN set & $N_{\rm SN}$ & $\ln B$ & $\Delta\chi^2_{\rm MAP}$ & $\Delta\mathrm{ELPD}_{\rm BAO}$ & $\Delta\mathrm{ELPD}_{\rm SN}$ & $\Delta\mathrm{ELPD}$ \\
\hline
Pantheon+ & 1590 & $-0.92 \pm 0.27$ & $-8.78$ & $-0.97$ & $+0.42$ & $-0.46$ \\
Pantheon+ (wide priors) & 1590 & $+2.36 \pm 0.30$ & $-8.33$ & $-0.98$ & $+0.45$ & $-0.44$ \\
DES-Y5 & 1829 & $-6.16 \pm 0.27$ & $-19.41$ & $-2.02$ & $-2.75$ & $-6.37$ \\
DES-Dovekie & 1820 & $-2.50 \pm 0.27$ & $-12.18$ & $-1.23$ & $+0.27$ & $-2.75$ \\
Union3 & 22 & - & $-15.01$ & $-1.79$ & $-1.34$ & $-3.83$ \\
\hline
\end{tabular}
\caption{Model comparison summary between $\Lambda$CDM and $w_0w_a$CDM cosmology models using DESI DR$2$ BAO, \textit{Planck} \texttt{CamSpec} and supernovae datasets. Every column is signed so that a negative value favours $w_0w_a$CDM and a positive value favours $\Lambda$CDM; $\Delta\chi^2_{\rm MAP} = \chi^2_{w_0w_a{\rm CDM}} - \chi^2_{\Lambda{\rm CDM}}$ (DESI sign), while $\ln B$ and $\Delta\mathrm{ELPD}$ are $\Lambda$CDM $-$ $w_0w_a$CDM. The two separate hold-out columns and the joint hold-out column are different cross-validation experiments, not a decomposition, and do not sum. Every row uses that dataset's default prior ranges except Pantheon+ (wide priors), which repeats the Pantheon+ fit on identical data and likelihoods with only the prior ranges widened (See Appendix~\ref{app:settings} for exact values); comparing those two rows isolates the prior sensitivity of each statistic. $\ln B$ is unavailable for Union3, whose chains were run with MCMC rather than nested sampling. Furthermore, the number corresponding to Union3 represents the number of Union3 bins, not the number of supernovae.}
\label{tab:results}
\end{table*}

In Table~\ref{tab:results}, we summarize the aggregated model selection metrics by probe, both for the separate hold-out case and the joint hold-out case. A simple global scalar number reading of each row gives the impression that $w_0w_a$CDM is preferred. However, a proper redshift decomposition of these results showcase the nuance of model predictive scoring. Since ELPD is a predictive density score, the significance of the numbers can be read as exponentiated odds ratio. For example, in Table~\ref{tab:results} row $1$, we see that the joint hold-out $\Delta$ ELPD is $-0.46$, meaning the odds ratio in favor of $w_0 w_a$CDM is $e^{0.46} \simeq 1.58:1$ against $\Lambda$CDM. We contextualize these results in the following subsections. 

\subsection{How the Choice of Priors Play A Role in Model Selection}
\label{sec:res-prior}

One key issue often noted in the cosmology community is that the Bayes Factor is extremely sensitive to prior, and a diffused prior choice will lead to a preference for $\Lambda$CDM \citep{cortes-liddle-bao-prior-2024, patel-prior-bao-2024}. As discussed in Section~\ref{sec:theory}, this is simply a consequence of Lindley's Paradox. Nevertheless, $\Delta$ ELPD is a Bayesian model-selection score that is insensitive to prior choices, provided the likelihood is contained within the prior volume.

We use the Pantheon+ dataset to demonstrate this behaviour, given that \cite{ong2026-handley-desi-dr2-bayesian} showed that with the official DESI prior volumes, the combination of Pantheon+, DESI BAO and \textit{Planck} CMB gives preference for $\Lambda$CDM under Bayes Factor, while $\Delta \chi^2_{\rm MAP}$ shows preference for $w_0w_a$CDM cosmology. We choose two sets of priors, a wide and a narrow prior (the default), as discussed in Appendix~\ref{app:settings}. For the dark energy equation of state, we chose tophat priors on $w_0-w_a$ with the additional condition that $w_0 + w_a < 0$.
 
 The resultant likelihoods for both prior choices are contained within the prior boundaries, and so the \textit{posteriors} that result are indistinguishable. As shown in the first two rows of Table~\ref{tab:results}, $\ln B$ shifts from $+2.36$ to $-0.92$ when we change the priors from wide to narrow, qualitatively changing the meaning of which model is preferred. On the other hand, the frequentist $\Delta\chi^2_{\rm MAP}$ metric moves by $0.45$, which we note is comparable to the scatter we obtain between minimizer restarts in the parameter inference. For the same setup, the $\Delta$ ELPD metric moves by $0.01$, $0.03$ and $0.02$ when considering separate hold-out of the BAO and supernovae data points, and the joint hold-out respectively. 

 In other words, $\Delta$ ELPD is practically insensitive to prior change, unlike $\ln B$, and it accounts for the performance of the full distribution (or chain), unlike $\Delta \chi^2_{\rm MAP}$, which is only a point-estimate. We stress that the $\Delta$ELPD score is not prior-independent by construction, the leave-one-out posteriors are broader than the full posterior, and a sufficiently tight prior would clip them, but that effect is measured here and is two orders of magnitude below the $\ln B$ shift.

We reiterate that in the current \textit{best practice} Bayesian in-sample model comparison method used regularly in cosmology and astrophysics, the role of the prior as an intrinsic model component is often overlooked. While the rationale may be to choose an ``uninformative" prior to maximize ignorance (by increasing prior volume), in practice this translates to the user indirectly saying that they trust the extreme value as much as the more ``reliable" values. A toy example can be thought of as the case where data is highly informative, leading the likelihood to approach a shrinking Gaussian-like distribution, while the prior remains unchanged. In such a scenario while the posterior will localize further and further, the absence of changing the prior essentially leads to that small posterior peak getting averaged out to $0$. 

\subsection{LRG2, not LRG1 or SNIa, Drives Dynamic Dark Energy Tension}

\begin{figure}[htbp!]
    \centering
    \includegraphics[width=\linewidth]{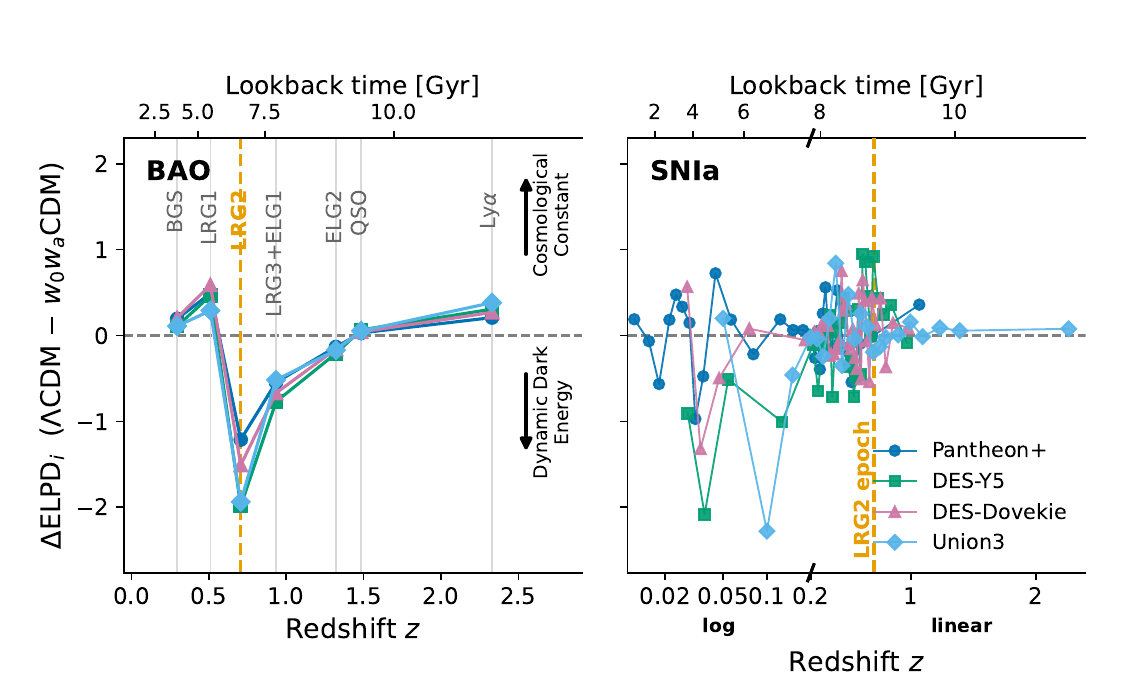}
    \caption{Per-redshift $\Delta$ELPD for the separate hold-out. (\textit{Left}) BAO predictive performance. Every compilation gives mild $\Lambda$CDM preferences at the lowest and highest redshifts and concentrates its $w_0w_a$CDM preference in a single block, LRG2. (\textit{Right}) Supernova predictive performance, which has no comparable feature: the per-block scores fluctuate about zero except for DES-Y5 at $z<0.1$, where calibration issues have been reported, and Union3, whose preference is concentrated in two bins. The contrast between the DES-Y5 and DES-Dovekie curves at low redshift is the effect of recalibration on the predictive score.}
    \label{fig:delta-elpd-sep}
\end{figure}

Using one scalar number to quantify model selection can lead to erroneously establishing a cause-and-effect relationship that may not be supported by data. For example, combining supernovae with DESI and CMB leads to the DESI + CMB posterior shrinking to the dynamic dark energy part of the $w_0w_a$ parameter space, positioning supernovae as the tracers that drive the dynamic dark energy tension. Fortunately, the $\Delta$ ELPD metric allows us to decompose the model predictive scoring across the redshift dimension to identify and quantify the most important points, which the in-sample tests like $\Delta \chi^2_{\rm MAP}$ and $\ln B$ are unable to do. We consider the predictive scoring when BAO and SN tracer data points are held out separately in Figure~\ref{fig:delta-elpd-sep}.

The left panel of Figure~\ref{fig:delta-elpd-sep} shows $\Delta\mathrm{ELPD}_i$ for the seven DESI DR2 BAO blocks. We see that the lowest and the highest two BAO redshift blocks -- BGS, LRG1, QSO and Ly$\alpha$ all favor $\Lambda$CDM mildly. 

Noticeably, the biggest evidence of $w_0w_a$CDM preference comes from two distinct blocks, and not coherently from across redshift ranges and tracers. For example, the value from LRG2, at $z = 0.706$, consistently scores between $-1.21$ and $-1.99$ depending on which supernova compilation occupies the conditioning set. Similarly, the LRG3+ELG1 block (the second most informative block) ranges between $-0.51$ and $-0.77$. However, the supernovae counterpart at the exact redshift of LRG2 ranges between $+0.92$ and $-0.19$, and for LRG3 the range is $+0.36$ and $-0.09$. This suggests that from a predictive scoring perspective, the $w_0w_a$CDM preference is \textit{entirely} driven by the LRG2 BAO point, and not the supernovae. If we remove only the LRG2 BAO point from the $\Delta$ ELPD scoring, we see that the overall score ranges between $+0.27$ and $-0.04$. If we compare this range with the overall scores in Table~\ref{tab:results}, we see that the qualitative preference for dynamic dark energy is entirely driven by LRG2. We note that \cite{ocolgain2026-desi-dr2} asked a separate but related question: ``the removal of which single BAO data point gives most consistency with $\Lambda$CDM?" They found LRG2 to be the biggest culprit in driving the tension. While an inference question cannot answer the problem of model selection, it is nevertheless worth noting that our out-of-sample model selection approach and their in-sample parameter inference approaches indicate that the problem lies with LRG2.

This result in contrast to the most \textit{statistically anomalous} point, which is LRG1. The anomalous nature of this point was first raised in the DR$1$ BAO analysis in a number of papers \citep{liu2024-lrg1-lgr2-dr1-anomaly, wang2024-lrg1-lrg2-dr1}, including the fact that it results in an unusually high $\Omega_{\rm M}$ in both DR$1$ \cite{ocolgain2026-dr1-anomaly} and in DR$2$ \cite{ocolgain2026-desi-dr2}. It may be surprising, therefore, that in the predictive scoring paradigm, the LRG1 point actually prefers $\Lambda$CDM. We note here that $\Delta\mathrm{ELPD}_i$ can address the question of \textit{which model predicts this block better than the other}: it says nothing about whether either prediction is good in an absolute sense. While beyond the scope of the current paper, we will address this problem in an upcoming companion paper that explores this question in the alternative basis of geometry and scale -- the Alcock-Paczy\'nski Factor, $F_{\rm AP}$ and the isotropic BAO scale, $D_V$, as well as alternatives to the CPL parametrization of dark energy using $\Delta$ ELPD. But as a preliminary result, we refer the readers to Appendix~\ref{app:1d-pred-density}, where we show that the hold-out predictions of $D_M/r_d$ and $D_H/r_d$ at LRG1 are better explained by models X and Y respectively, but also that the actual data lives in the tail of these predictions. Thus, while LRG1 is the most anomalous point, \textit{both} $\Lambda$CDM and $w_0w_a$CDM fail to predict this point the same way, leading to its lack of impact in model selection.

\begin{figure}[htbp!]
    \centering
    \includegraphics[width=0.9\linewidth]{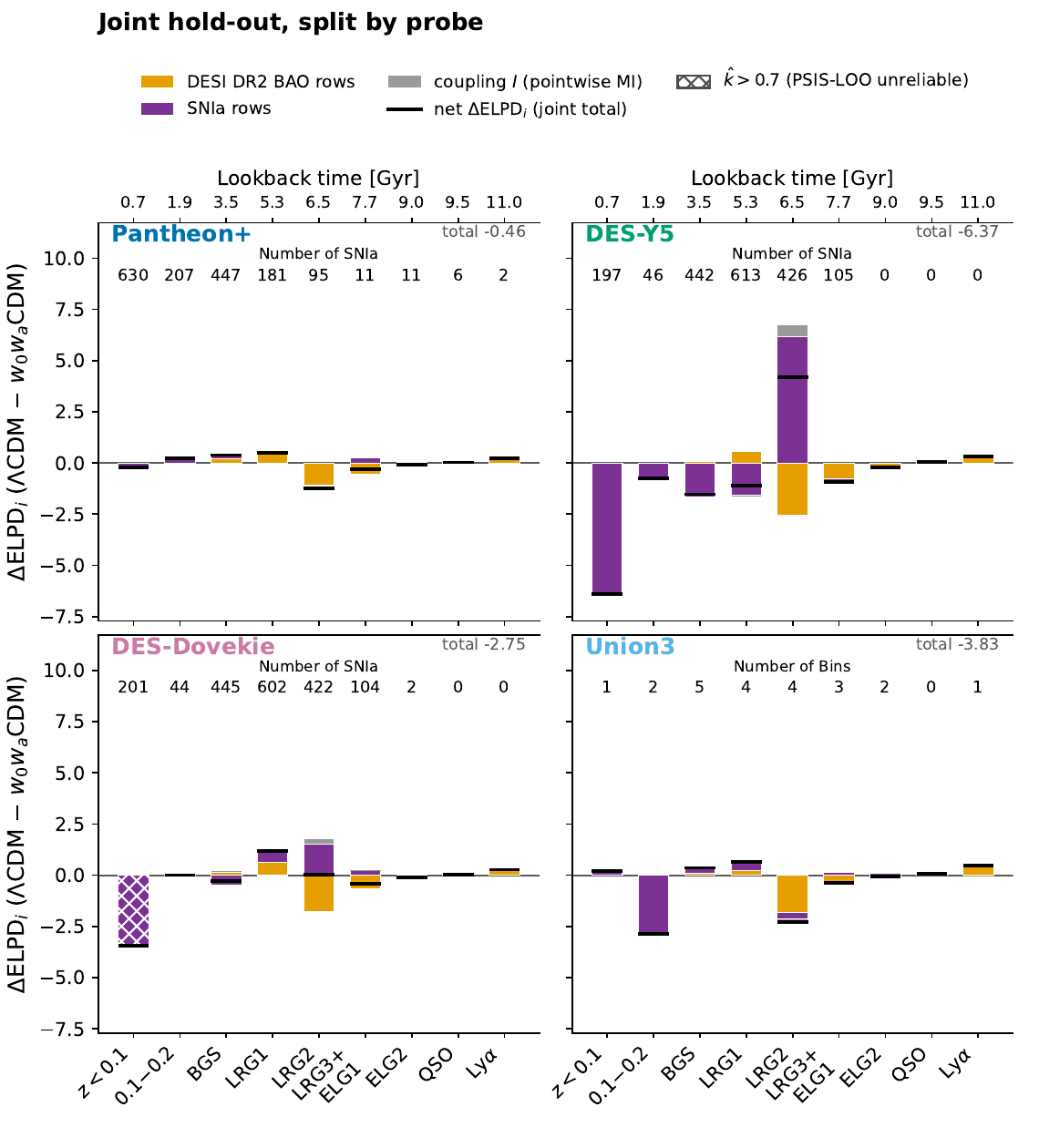}
    \caption{Joint hold-out: each redshift cell is withheld in full --- the BAO block together with every supernova whose $z_{\rm cmb}$ falls in the cell --- and the resulting $\Delta$ELPD is decomposed into its BAO rows, its supernova rows, and the coupling term $I$. Panels show the four supernova compilations. The number of withheld observables is printed beneath every bar, because $\Delta$ELPD is extensive and bar heights from cells of very different size are not comparable. Blocks with $\hat{k}$ above the reliability threshold are marked.}
    \label{fig:delta-elpd-joint}
\end{figure}

We see a similar story emerge when we compare separate hold-out versus joint hold-out analyses; the joint hold-out analysis is shown in Figure~\ref{fig:delta-elpd-joint}. In the joint hold-out analysis, we see how a model makes predictions when no longer can access complementary information from the supernovae (or the BAO) to help the BAO (or the supernovae). A joint prediction leads to the posterior part of Equation~\ref{eq:loo-predictive} to change from the separate hold-out case because in this scenario the posterior is not $p(\theta | y_{\rm BAO, seen}, y_{\rm SN})$ but rather $p(\theta | y_{\rm BAO, seen}, y_{\rm SN, seen})$. However, since the likelihood is still decomposable, we can compute what amount of the joint $\Delta$ELPD is driven by BAO exclusively, by the supernovae exclusively, and the mutual information that describes how the two tracers can in principle co-vary; this last term is represented as the coupling term $I$ in the figure). 

We see that for Pantheon+, the evidence for either side is modest, with most of the preference for $w_0w_a$ again coming from LRG2. In contrast, DES-Y5 and DES-Dovekie comparisons show interesting results. DES-Y5 overall shows stronger preference for $w_0w_a$CDM model below and including LRG1, but at LRG2 the BAO point solely prefers dynamic dark energy while the supernovae in the same redshift overwhelmingly prefer $\Lambda$CDM. This support for dynamic dark energy almost disappears in DES-Dovekie except at the lowest redshift bin. We discuss this difference further in Section~\ref{sec:sn-results}. The net result at LRG2 in DES-Dovekie may suggest that neither of the models predict better than the other, however the decomposition shows that supernovae and the BAO point in different directions by almost the same amplitude, thus cancelling out their effects. In Union3, we also see that except the $0.1 \leq z \leq 0.2$ bin, LRG2 is the only block that again shows preference for dynamic dark energy. The consistent picture that emerges in both separate-hold-out and joint-hold-out is  that supernovae on their own have very little predictive preference for dynamic dark energy, and in the BAO sector it is entirely LRG2-driven. 

\subsection{Rediscovering the $z < 0.1$ SN Miscalibration via Predictive Scoring}
\label{sec:sn-results}
The joint ELPD total for the DES-Y5 compilation is $-6.40$ is in the $z<0.1$ cell alone. There is no BAO measurement in this cell, so here the entire contribution is from supernovae. The DES-Dovekie recalibration reduces the same cell to $-3.46$, and Pantheon+ and Union3 show nothing comparable. This is an out-of-sample recovery of the low-redshift magnitude offset identified by \citet{desy5-syst-efstathiou2025} and discussed by \citet{desy5-syst-huang2025}, and it is not incidental to the method. Because we analytically marginalize over the absolute magnitude $M_B$ (following the \textsc{cobaya} convention), the supernova likelihood constrains only relative distance moduli, and withholding a redshift range of supernovae and predicting it \textit{is structurally} a test of that range's calibration relative to the rest of the sample. The result is therefore reached without inspecting an individual supernova, without comparing compilations object by object, and without any model of the systematic. This result is more clearly visible in the separate hold-out analysis in Figure~\ref{fig:delta-elpd-sep}. For points at $z < 0.2$, we see that while DES-Y5 likelihood drives preference for dynamic dark energy, the recalibrated DES-Dovekie in the same range dilutes that preference. Thus, we also discover in a predictive sense how much the $w_0w_a$ parametrization soaks up the color calibration systematic as a signature of dark energy. 

Our analysis for the low-$z$ supernovae has one crucial caveat. The $z < 0.1$ cell carries the largest Pareto $\hat{k}$ in the analysis, $0.75$ for DES-Dovekie, so the importance-sampling estimate is at or beyond its reliability threshold there. A large $\hat{k}$ is itself a statement that the remaining data predict this block poorly. In other words, the $z < 0.1$ data in DES-Dovekie is significantly different enough that the PSIS-LOO reweighting technique will lead to biased ELPD scores, and one has to re-run the chain by explicitly removing these supernovae to compute ELPD and use beyond-LOO-CV algorithms to estimate ELPD \citep{elpd-bregman-divergence}.

\section{Discussion and Conclusion}
\label{sec:discussion}

We used leave-one-redshift-block-out cross-validation estimator of the expected log predictive density model selection metric to address the significance of the dynamic dark energy signature in the DESI DR$2$ BAO analysis, and identify which redshifts drive this tension. We find that:
\begin{enumerate}
\item \textbf{$\Delta$ELPD is insensitive to the prior } On indistinguishable posteriors, $\ln B$ changes sign, thereby switching votes between $w_0w_a$CDM and $\Lambda$CDM cosmology from the data, while $\Delta$ELPD moves by $0.02$. In contrast, while $\Delta \chi^2_{\rm MAP}$ is also insensitive to prior, it is however unable to account for the full distribution, as it is strictly a point-estimate. $\Delta$ ELPD accounts for the behavior of the full distribution in comparison. 

\item \textbf{The out-of-sample preference for $w_0w_a$CDM is modest in aggregate and highly localized in detail.} In both separate-hold-out and joint-hold-out analyses, the strongest preference for dynamic dark energy comes from DES-Y5 chains. In contrast, Pantheon+ shows the least preference, in fact showing marginal preference for $\Lambda$CDM when looking at the $\Delta$ ELPD of supernovae exclusively.

\item \textbf{One BAO redshift block supplies the entire BAO preference.} LRG2 ($z = 0.706$) consistently drives the preference for dynamic dark energy, irrespective of separate versus joint-hold-out analyses, or supernovae dataset. The strength of the $w_0w_a$CDM preference in aggregate is entirely driven by this one BAO point; removing LRG2 swings the model selection metric's preference back to $\Lambda$CDM. However, the most anomalous point, LRG1, plays little role in predictive model selection because its measured value cannot be explained by either $\Lambda$CDM or $w_0w_a$CDM in a predictive sense. The lowest redshift BAO point, BGS, is highly consistent with $\Lambda$CDM predictiions; this is the redshift where the signature of dark energy should be the strongest.                                    
\item \textbf{The DES-Y5 preference is a low-redshift calibration statement, and is derived rather than assumed.} It concentrates at $z<0.1$ and the preference for dynamic dark energy reduces from $-6.40$ to $-3.46$ under the DES-Dovekie recalibration. Because $M_B$ is marginalized, a supernova redshift-block hold-out is structurally a relative-calibration test, so this is an independent derivation of the offset reported by \citet{desy5-syst-efstathiou2025}.
\end{enumerate}

We also note that the reader may have questions regarding the uncertainty of the ELPD measurements, especially per redshift block. We show in Equation~\ref{eq:loo-predictive} that while the likelihood part is deterministic for an experiment, the posterior part depends on the sampler. In principle, this sampler noise could contribute to the ELPD estimator. We checked with $100$ bootstrap realizations per experiment and chains that the sampler noise is only about $1\%$ of the $\Delta$ELPD values. Another source of error could be from the estimation process using leave-one-redshift-block-out cross-validation; we are restricted to only $7$ BAO blocks with the current DESI configuration. One could do a jack-knife style computation of the overall $\Delta$ ELPD; however, we caution against that given that a significant point like LRG2 can completely dominate the overall analysis. Thus, our recommendation for future BAO analyses is to publish results in a way that allows others to re-bin the data (rather than in a highly compressed likelihood), to allow proper re-sampling, in a similar fashion to supernovae likelihoods and datasets.

\noindent While several analyses focus on in-sample metrics for model comparison, we use out-of-sample metrics to robustly show that the case for evolving dark energy in DESI DR2 rests, on one BAO measurement at $z=0.706$. In an upcoming paper, we aim to explore the physical basis for why LRG2 is the driver of the tension in $F_{\rm AP}$ and $D_V/r_d$ bases, along with alternatives to CPL parametrization and also suggest principled approaches to designing priors for Bayesian in-sample analysis.

\begin{acknowledgments}
TK was funded by the Dunlap Institute Postdoctoral Fellowship. The Dunlap Institute is funded through an endowment established by the David Dunlap family and the University of Toronto. All the computation of this research was done on the Digital Research Alliance of Canada Trillium cluster. The authors used \textsc{Claude}-\texttt{Opus} and \texttt{Sonnet} Large Language Models to help improve the visualizations.

JSS was supported by NSERC Discovery Grant RGPIN-2023-04849. RH also acknowledges support from the NSERC Discovery Grant Program RGPIN-2025-06483 and the Arthur B. McDonald Fund SMFSU-60768. The authors at the University of Toronto acknowledge that the land on which the University of Toronto is built is the traditional territory of the Wendat Nation, the Seneca, and the Mississaugas of the Credit. Today, this meeting place is still the home to many Indigenous people from across Turtle Island. The authors are grateful to have the opportunity to work on this land. 

\end{acknowledgments}

\software{\textsc{cobaya} \citep{cobaya2012, cobaya2019}, \textsc{camb} \citep{camb2000, camb2012}, \textsc{polychord} \citep{polychord2015}, \textsc{anesthetic} \citep{anesthetic}, \textsc{arviz}, \textsc{numpy}, \textsc{scipy}, \textsc{getdist}, \textsc{matplotlib}, \textsc{Claude}-\texttt{Opus} and \textsc{Claude}-\texttt{Sonnet}.}


\bibliography{sample701}{}
\bibliographystyle{aasjournalv7}

\appendix

\section{Sampler, Theory-Code, and Prior Settings}
\label{app:settings}

\subsection{Sampling}

\begin{table}[h!]
\centering
\begin{tabular}{llcc}
\hline\hline
 & Parameter & Baseline & Wide \\
\hline
\multirow{6}{*}{Shared}
 & $H_0$ [km\,s$^{-1}$\,Mpc$^{-1}$] & $[60,\ 80]$ & $[50,\ 90]$ \\
 & $\ln(10^{10} A_s)$              & $[2.9,\ 3.13]$ & $[1.8,\ 4.0]$ \\
 & $\Omega_{\rm c} h^2$            & $[0.10,\ 0.13]$ & $[0.01,\ 0.30]$ \\
 & $\Omega_{\rm b} h^2$            & $[0.02,\ 0.03]$ & $[0.01,\ 0.04]$ \\
 & $n_s$                           & $[0.94,\ 1.0]$ & $[0.85,\ 1.1]$ \\
 & $\tau$                          & $[0.02,\ 0.10]$ & $[0.01,\ 0.15]$ \\
\hline
\multirow{2}{*}{$w_0w_a$CDM only}
 & $w_0$ & $[-1.11,\ -0.57]$ & $[-3,\ 1]$ \\
 & $w_a$ & $[-1.57,\ 0.46]$ & $[-3,\ 2]$ \\
\hline
 & $(w_0, w_a)$ area & $1.096$ & $15.5$ \\
\hline
\multirow{2}{*}{Fixed}
 & $\sum m_\nu$ [eV] & \multicolumn{2}{c}{$0.06$ (one massive eigenstate)} \\
 & $N_{\rm eff}$ & \multicolumn{2}{c}{$3.046$} \\
\hline
\multirow{2}{*}{Nuisance}
 & $A_{\rm planck}$, $\mathrm{amp}_{143}$, $\mathrm{amp}_{217}$,
   $\mathrm{amp}_{143\times217}$, & \multicolumn{2}{c}{\multirow{2}{*}{\texttt{cobaya} defaults}} \\
 & $n_{143}$, $n_{217}$, $n_{143\times217}$, $c_{TE}$, $c_{EE}$ & & \\
\hline\hline
\end{tabular}
\caption{Sampled parameters and prior ranges. All priors are uniform over the stated interval, and we impose $w_0 + w_a < 0$ in both configurations; the quoted $(w_0, w_a)$ areas are those of the allowed region after this constraint. The nine nuisance parameters of the CamSpec and lensing likelihoods are common to both models with identical priors and therefore cancel in the Bayes factor. Sampled dimensionalities are $n_{\rm dim} = 15$ ($\Lambda$CDM) and $17$ ($w_0w_a$CDM). Each baseline range spans at least $5\sigma$ of the corresponding marginal posterior, so the likelihood is contained in both configurations (Appendix~\ref{app:occam-check}). The baseline ranges are posterior-informed and are used for the ELPD analysis, where prior width is demonstrably irrelevant; see Section~\ref{sec:sampling} for the caveat attached to Bayes factors computed under them.}
\label{tab:priors}
\end{table}

We sample the posterior with the \textsc{cobaya} interface to \textsc{polychord} \citep{polychord2015} for the Pantheon+, DES-Y5 and DES-Dovekie combinations, and with the \textsc{cobaya} \citep{cobaya2012, cobaya2019} Metropolis sampler for Union3. The sampled parameter space comprises the six $\Lambda$CDM parameters (eight for $w_0w_a$CDM) together with nine nuisance parameters common to both models --- the overall calibration $A_{\rm planck}$, six \texttt{CamSpec} foreground amplitudes and spectral indices ($\mathrm{amp}_{143}$, $\mathrm{amp}_{217}$, $\mathrm{amp}_{143\times217}$, $n_{143}$, $n_{217}$, $n_{143\times217}$), and the polarization calibrations $c_{TE}$ and $c_{EE}$ --- giving $n_{\rm dim} = 15$ and $17$ respectively. Because these nine enter both models with identical priors, they cancel exactly in the Bayes factor.

For \textsc{polychord} we use \texttt{nlive} $= 1000$, \texttt{nprior} $=1500$, \texttt{num\_repeats} $= 3 n_{\rm dim}$ (45 and 51), and a precision criterion of $0.01$. The live-point count exceeds the usual $25 n_{\rm dim}$ guidance for evidence estimation by a factor of $2.4$--$2.7$. Evidences are computed using \texttt{anesthetic} \citep{anesthetic}. For the Metropolis sampler we use \texttt{proposal\_scale} $= 1.5$, \texttt{Rminus1\_stop} $= 0.01$, \texttt{Rminus1\_cl\_stop} $= 0.2$, \texttt{drag} $=$ \texttt{True}, \texttt{learn\_proposal} $=$ \texttt{True}, and \texttt{max\_tries} $= 5000$.

While both samplers return weighted output, we need to pass the equal-weighted chains to \texttt{arviz.loo} for the ELPD computation. We use the package \textsc{anesthetic} to convert the native chains into equal-weighted chains. 

\subsection{Theory Code and Cosmological Assumptions}

We use \textsc{camb} \citep{camb2000, camb2012} via \textsc{cobaya} \citep{cobaya2012, cobaya2019}, with the PPF dark energy implementation for $w_0w_a$CDM so that $w(z)$ may cross $-1$. For the non-linear matter power spectrum entering the CMB lensing likelihood we adopt \texttt{mead2016}, and we set \texttt{lens\_potential\_accuracy} $= 4$, \texttt{lmax} $= 4000$ and \texttt{lens\_margin} $= 1250$; the accuracy parameters \texttt{AccuracyBoost}, \texttt{lAccuracyBoost} and \texttt{lSampleBoost} are left at their defaults of unity. We fix $\sum m_\nu = 0.06$~eV in a single massive eigenstate and $N_{\rm eff} = 3.046$, matching \citet{ong2026-handley-desi-dr2-bayesian} so that our Bayes factors are directly comparable to theirs. (The currently preferred value is $3.044$; the difference is cosmologically negligible here.)

\subsection{Chains and Priors}
\label{sec:sampling}

Table~\ref{tab:priors} lists the sampled parameters and the two prior choices we compare. We impose $w_0 + w_a < 0$ throughout. Our baseline runs use priors restricted to regions the data already occupy, which is computationally efficient and legitimate \textit{precisely} because ELPD is insensitive to prior width. We stress that this narrow prior is posterior-informed and is therefore not a defensible \textit{a priori} prior for an evidence calculation; we report $\ln B$ under it in Table~\ref{tab:results} for completeness and for comparison across compilations. Note that our Bayes Factor as a result will not match published results owing to differences in priors. 

\pagebreak
\section{Posterior Predictive of Held-Out Data Conditioned on Seen Data}
\label{app:1d-pred-density}

\begin{figure}[htbp!]
    \centering
    \includegraphics[width=0.77\linewidth]{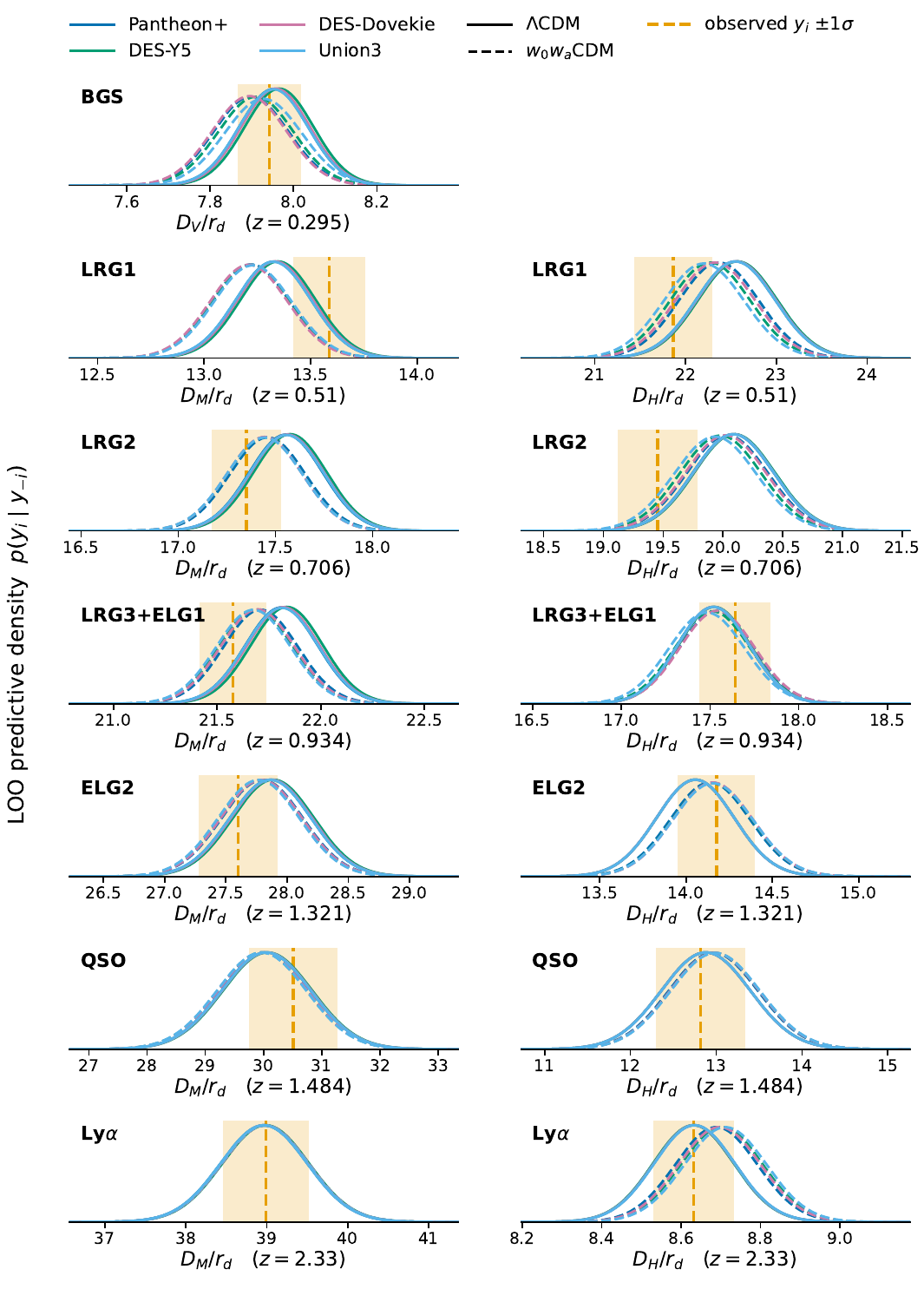}
    \caption{$1$D predictive density of DESI DR$2$ BAO points as a function of cosmological models and supernovae datasets. BGS only has $D_V/r_d$ measurement, while the other points have $D_M/r_d$ and $D_V/r_d$ measurements. At higher redshifts and at BGS, $\Lambda$CDM predicts observed data well. LRG1 and LRG2 points are the most discordant with predictions, with $\Lambda$CDM prediction of LRG1 $D_M/r_d$ and $w_0w_a$CDM prediction of LRG1 $D_H/r_d$ being better than the alternate respectively. For LRG2, $w_0w_a$CDM gives a better prediction along $D_M/r_d$ axis, while both models perform poorly along the $D_H/r_d$ axis.}
\end{figure}

\end{document}

%% file: bangla_commands.tex
\def\bng{\bngx}

\font\bngx=bang10

\def\*#1*#2{o\null{#2}{#1}}

\def\d#1{\oalign{\smash{#1}\crcr\hidewidth{$\!$\rm.}\hidewidth}}

\def\sh#1{\setbox0=\hbox{#1}%
     \kern-.02em\copy0\kern-\wd0
     \kern.04em\copy0\kern-\wd0
     \kern-.02em\raise.0433em\box0 }